\documentclass[a4paper,11pt]{article}
\usepackage{jcappub} % for details on the use of the package, please
\usepackage[T1]{fontenc} % if needed
\RequirePackage{bm,amsfonts,lineno,amsmath,amssymb,float,eurosym,latexsym,epsf,mathtools,cuted,makecell,array}
\usepackage[font={footnotesize,it}]{caption}
\setcellgapes{4pt}%parameter for the spacing

\title{\boldmath Cosmological Analysis of $f(R, \Sigma, T)$ Gravity with EoS Parameterization}

\author[a,b,1]{S. H. Shekh\note{Corresponding author.},}
\author[b]{N. Myrzakulov,}
\author[c]{Anil Kumar Yadav,}
\author[d]{Anirudh Pradhan}

\affiliation[a]{Department of Mathematics, S.P.M. Science and Gilani Arts, Commerce College, Ghatanji, Yavatmal, Maharashtra-445301, India.}
\affiliation[b]{L N Gumilyov Eurasian National University, Astana 010008, Kazakhstan}
\affiliation[c]{Department of Physics, United College of Engineering and Research,Greater Noida - 201310, India}
\affiliation[d]{Centre for Cosmology, Astrophysics and Space Science, GLA University, Mathura-281 406,\\ Uttar Pradesh, India}

\emailAdd{da\_salim@rediff.com}
\emailAdd{nmyrzakulov@gmail.com}
\emailAdd{abanilyadav@yahoo.co.in}
\emailAdd{pradhan.anirudh@gmail.com}

\abstract{In this paper, we present a comprehensive cosmological analysis within the framework of $f(R, \Sigma, T)$ gravity, a modified theory that incorporates nonlinear matter-geometry coupling via the inclusion of both the trace of the energy-momentum tensor $T$ and the scalar $\Sigma = T_{\mu\nu}T^{\mu\nu}$. We consider a spatially flat Friedmann-Robertson-Walker (FRW) universe and introduce a linear parameterization for the equation of state (EoS) parameter based on the Chevallier-Polarski-Linder (CPL) form, which allows us to explore the dynamical evolution of dark energy without imposing restrictive assumptions. To confront the theoretical model with observations, we utilize the latest Hubble parameter measurements from cosmic chronometers. The model parameters are constrained using Markov Chain Monte Carlo (MCMC) simulations with the \texttt{emcee} package, leading to tight bounds on the parameters. The analysis reveals that the model remains consistent with the $\Lambda$CDM scenario while allowing mild deviations consistent with observational data. Furthermore, we examine the physical and kinematical features of the model by studying the behavior of the physical parameters. Finally, the calculated age of the universe within this framework is found to be in excellent agreement with Planck 2018 results, highlighting the and viability of $f(R, \Sigma, T)$ gravity as a candidate for explaining late-time cosmic acceleration. }


\begin{document}
\maketitle
\flushbottom

\section{Introduction}
	
	The accelerated expansion of the Universe, as revealed by several independent observational probes such as \textit{type Ia supernovae} (SNe Ia) \cite{Riess1998, Perlmutter1999}, \textit{cosmic microwave background anisotropies} (CMB) \cite{Planck2018}, and \textit{baryon acoustic oscillations} (BAO) \cite{Eisenstein2005}, has posed significant challenges to the standard model of cosmology based on General Relativity (GR) and the cosmological constant $\Lambda$. The simplest explanation within GR is the inclusion of a dark energy component with negative pressure; however, this approach suffers from theoretical difficulties such as fine-tuning and coincidence problems \cite{Weinberg1989, Sahni2000, Peebles2003}. This has motivated researchers to explore modified theories of gravity as a plausible alternative to account for the late-time cosmic acceleration without invoking unknown exotic components such as dark energy or dark matter.

In recent years, various extensions of General Relativity have been extensively explored to address the limitations of the standard cosmological model. Among these, $f(R)$ gravity \cite{Sotiriou2010, DeFelice2010, Nojiri2011} remains a foundational modification wherein the Ricci scalar $R$ is generalized to an arbitrary function, offering richer cosmic dynamics. Another important class is $f(T)$ gravity, where $T$ denotes the torsion scalar, providing a viable alternative through teleparallel geometry \cite{Cai2016}. The combination of curvature and torsion has also led to hybrid models such as $f(T,B)$ gravity, with $B$ representing the boundary term \cite{Bahamonde2015}. Furthermore, matter-coupled gravities like $f(R,T)$ \cite{Harko2011, Moraes2017} and $f(Q,T)$ theories, where $Q$ is the non-metricity scalar from symmetric teleparallel gravity, allow direct interaction between matter and geometry, introducing novel features in cosmic evolution \cite{Xu2023, Mandal2020}. The $f(Q)$ framework, in particular, has attracted attention for its second-order field equations and capability to describe late-time acceleration without invoking dark energy \cite{Jimenez2018, Lazkoz2019}. These modified gravity models not only offer flexibility in explaining observational data but also provide platforms to investigate the fundamental nature of gravitational interaction and the unification of cosmic phenomena.	 Recently, $f(R,T)$ framework has been extended even further to include dependence on $\Sigma = T_{\mu\nu} T^{\mu\nu}$, resulting in the so-called $f(R, \Sigma, T)$ gravity \cite{Xu2019, Sharif2022, Moraes2022}. The inclusion of $\Sigma$ brings nonlinear matter coupling effects that may contribute to explaining both early and late-time acceleration, galaxy rotation curves, and other astrophysical phenomena without dark matter or dark energy hypotheses.
	
	Parameterizing the equation of state (EoS) of the cosmic fluid allows a model-independent way to study the nature of cosmic acceleration and its evolution \cite{Chevallier2001, Linder2003, Barboza2008}. Several EoS parameterizations have been proposed and successfully confronted with observations to constrain cosmological parameters \cite{Zhai2013, Wang2018}. The combination of $f(R, \Sigma, T)$ gravity with EoS parameterizations thus provides a fertile ground to investigate the dynamical behavior of the Universe and confront theoretical predictions with observational data.
	
	In the present work, we consider the $f(R, \Sigma, T)$ gravity framework and adopt a suitable EoS parameterization to derive the modified Friedmann equations and obtain the expression for the Hubble parameter $H(z)$. We then utilize observational datasets to estimate and constrain the model parameters, thereby testing the viability of the model in light of current cosmological observations. The motivation behind choosing $f(R, \Sigma, T)$ gravity lies in its ability to incorporate more generalized matter-geometry couplings, offering a richer structure to study both early and late-time cosmic dynamics. This extended gravity model allows us to explore new physical effects emerging from the nonlinear interaction between matter and geometry, which are absent in standard $f(R)$ or $f(R, T)$ theories.
		The action principal to derive a set of field equations is defined as
	\begin{small}
		\begin{equation}\label{1}
			I = \frac{1}{16\pi} \int_{\Omega} \sqrt{-g} \left( B + \mathcal{L}_m \right) d^4x = \frac{1}{16\pi} \int_{\Omega} \sqrt{-g} \left( g^{ab} B_{ab} + \mathcal{L}_m \right) d^4x.
		\end{equation}
	\end{small}
	where $\Omega$ is a region enclosed within some closed surface, $B_{ab}=R_{xy}\left\{ \right\}+\Sigma (\Sigma_{xy})$ and $\Sigma_{xy} = b \Phi(xy)$, $b$ be the parameter which is the ratio between gravity and antigravity (i.e. attraction and repulsion) within a given system. The action principle can be expressed in the following manner
	
	\begin{equation}\label{2}
		I = \frac{1}{16\pi} \int_{\Omega} \sqrt{-g} \left( f(R, \Sigma, T) + \mathcal{L}_m \right) d^4x.
	\end{equation}
	
	By varying the gravitational action (\ref{2}) with respect to the metric tensor, we can arrive at the field equation for $f(R, \Sigma, T)$  Gravity as follows
	%\begin{widetext}
		\begin{equation}\label{3}
			R_{ab} \frac{\partial f}{\partial R} + R_{ab} \frac{\partial f}{\partial R} - \frac{1}{2} g_{ab} f + \left( g_{ab} \nabla^\gamma \nabla_\gamma - \nabla_a \nabla_b \right) \frac{\partial f}{\partial R} - \beta \frac{\partial f}{\partial R} \\
			= 8\pi T_{ab} +\left( T_{ab} + p g_{ab}\right) \frac{\partial f}{\partial T},
		\end{equation}
	%\end{widetext}
	The energy–momentum tensor for matter is defined as
	\begin{equation}\label{4}
		T_{ab} = (\rho + p) u_a u_b -  g_{ab} p,
	\end{equation}
	%	\begin{equation}\label{e4}
		%		T_{ab} = (\rho + p_{eff}) u_a u_b -  g_{ab} p_{eff},
		%	\end{equation}
	where $\rho$ is the energy density of matter, $p$ is isotropic pressure %and is given by $p_{eff} =$ matter pressure $(p)$+ viscous pressure ($\Pi = -3H\xi$) and $\xi=\xi_0+\xi_1H+\xi_2(\dot{H}+H^2)$; $\xi_0, \xi_1, \xi_2$ are positive constants. 
	$u_a$ is the fluid-four velocity vector, where $u_a u^b =1$. In this article, we assume that the function $f(R, \Sigma, T)$ is
	given by
	\begin{equation}\label{5}
		f(R, \Sigma, T) = R + \Sigma + 2\pi \eta T.
	\end{equation}
	where $\eta$ is an arbitrary constant parameter. Using the function (\ref{5}), the final form of $f(R, \Sigma, T)$ equation (\ref{3}) is given by
	\begin{small}
		\begin{equation}\label{6}
			R_{ab} + \Sigma_{ab} - \frac{1}{2} g_{ab} (R + \Sigma) = 2(4\pi + \pi \eta) T_{ab} + \pi \eta g_{ab}(T + 2p)
		\end{equation}
	\end{small}
	which can be alternatively expressed as
	\begin{small}
		\begin{equation}\label{7}
			B_{ab} = R_{ab} + \Sigma_{ab} = 2(4\pi + \pi \eta) \left(T_{ab} - \frac{1}{2} g_{ab} T\right) - \pi \eta g_{ab} (T + 2p).
		\end{equation}
	\end{small}

	Having established the general form of the field equations for $f(R, \Sigma, T)$ gravity in equation (\ref{7}), we now proceed to investigate the cosmological implications of this framework. In order to explore the dynamical evolution of the Universe under this modified gravity theory, it is essential to consider a specific spacetime geometry compatible with the large-scale homogeneity and isotropy observed in the Universe. For this purpose, we adopt the spatially flat Friedmann-Robertson-Walker (FRW) metric, which serves as a suitable cosmological background. Utilizing this metric, the field equations can be reduced to a set of modified Friedmann equations, allowing us to derive explicit expressions for key cosmological quantities such as the Hubble parameter, energy density, and pressure. These expressions will form the foundation for the subsequent analysis of the equation of state parameter and its role in driving the cosmic acceleration.

%\textcolor{red}{The article is organized as follows: In Section II, we present the basic formalism of $f(R, \Sigma, T)$ gravity. In Section III, we introduce the EoS parameterization and derive the corresponding expression of the Hubble parameter. Section IV is devoted to the estimation and constraint of model parameters using observational datasets. Finally, in Section V, we summarize our results and discuss the physical implications of the obtained constraints.}

	%%%%%%%%%%%%%%%%%%%%%%%%%%%%%%%%%%%%%%%%%%%%

%%%%%%%%%%%%%%%%%%%%%%%%%%%%%%%%%%%%%%%%%%%%%%%%%%%%%
	\section{Metric and Field Equations in $f(R, \Sigma, T)$ Gravity}
	
To explore the cosmological implications of the $f(R,\Sigma,T)$ gravity framework, we adopt a general form for the metric and specialize it for a spatially homogeneous and isotropic Friedmann-Robertson-Walker (FRW) space-time.
	\begin{equation}\label{8}
		ds^{2}=-dt^{2}+\delta_{ij} g_{ij} dx^{i} dx^{j}, \quad i,j=1,2,3,\dots,N,
	\end{equation}
	where $g_{ij}$ represents the spatial metric components, which are functions of the cosmic time $t$ and spatial coordinates $x^1, x^2, x^3$. Here, $t$ is measured in gigayears (Gyr), aligning with the cosmological time convention. Utilizing this metric within a comoving coordinate system allows us to derive the fundamental field equations for $f(R,\Sigma,T)$ gravity from the generalized field equations. For a spatially homogeneous and isotropic universe, the metric components in a four-dimensional Friedmann –Robertson–Walker (FRW) space-time simplify as:
	\begin{equation}\label{9}
		\delta_{ij} g_{ij} = a^{2}(t),
	\end{equation} 
	where $a(t)$ denotes the scale factor of the universe, encapsulating the overall expansion dynamics. In this FRW background, the metric components satisfy the condition $g_{11} = g_{22} = g_{33} = a^{2}(t)$, illustrating the equivalence among spatial directions.
	
	By employing the metric in Eq. (\ref{9}) within a comoving coordinate system, we derive the fundamental field equations for $f(R, \Sigma, T)$ gravity from the generalized field equations:
	\begin{equation}\label{10}
		3(b-1)\left(\dot{H} + H^2\right) = 4 \pi \rho + 4 \pi (3+\eta) p,
	\end{equation}
	\begin{equation}\label{11}
		3(1-b)^2 H^2 = \pi (8+3\eta) \rho - \pi \eta p,
	\end{equation}
		where $H = \dot{a}/a$ is the Hubble parameter, $b$ is a model parameter, and $\eta$ encapsulates modifications due to the extended gravity framework. 
		
	Recent investigation on the $f(R,\Sigma,T)$ gravity model was performed in \cite{New2025PLB}, where the authors explored exact cosmological solutions in the presence of nonlinear matter couplings. Their study demonstrated that the inclusion of both $T$ and $\Sigma$ terms significantly modifies the behavior of cosmic evolution, allowing the model to accommodate both decelerated and accelerated phases without invoking additional exotic fields. The authors also emphasized that the interaction between $\Sigma$ and the geometry can naturally generate a transition from matter-dominated to dark energy-dominated epochs, consistent with late-time cosmic acceleration. These findings further support the viability of the $f(R,\Sigma,T)$ framework and provide complementary evidence to the present work, where we adopt a more general parameterization via the equation of state to analyze observational consistency. From the modified field equations (\ref{10}) and (\ref{11}), one can observe that the cosmic evolution in $f(R, \Sigma, T)$ gravity depends on the parameters $b$ and $\eta$, along with the energy density $\rho$ and pressure $p$ of the cosmic fluid. To proceed further, we express these two equations in such a way that we can isolate $p$ and $\rho$ explicitly in terms of the Hubble parameter $H$ and its time derivative $\dot{H}$.
	
	Starting from equation (\ref{10}), we can solve for the combination involving $H$ and $\dot{H}$:
	\begin{equation}\label{12}
		p = \frac{(1-b)}{4(8\pi + 6\eta + \pi \eta^2)} \left[ (8 + 3\eta)\dot{H} + (4(3-b) + 3\eta) H^2 \right].
	\end{equation}
	
	Similarly, from equation (\ref{11}), we can express $\rho$ in terms of $H$ as
	\begin{equation}\label{13}
		\rho = \frac{(1-b)}{4(8\pi^2 + 6\pi\eta + \eta^2)} \left[ -\eta \dot{H} + (12\pi(1-b) + \eta(3-4b)) H^2 \right].
	\end{equation}
	
 	The above two equations form a system of two linear equations in $\rho$ and $p$. Having obtained the explicit forms of $p$ and $\rho$, we can now define the equation of state (EoS) parameter $\omega$, which characterizes the nature of the cosmic fluid and governs the expansion dynamics of the Universe. The EoS parameter is defined as
	\begin{equation}\label{14}
		\omega = \frac{p}{\rho}.
	\end{equation}
	
	Substituting the expressions from equations (\ref{12}) and (\ref{13}) into equation (\ref{14}), we obtain the explicit expression for $\omega$ as
	\begin{equation}\label{15}
		\omega = - \frac{(\eta + 2\pi)(\eta + 4\pi)\left[ H^2(3(\eta+4) - 4b) + (3\eta + 8)\dot{H} \right]}{\left(\pi (\eta^2 + 8) + 6\eta\right) \left[ H^2 \left((4b - 3)\eta + 12\pi(b-1)\right) + \eta \dot{H} \right]}.
	\end{equation}
	
	This relation demonstrates how the EoS parameter depends on both the Hubble parameter and its derivative, as well as on the coupling parameters $b$ and $\eta$ of the $f(R,\Sigma,T)$ gravity model. The evolution of $\omega$ provides crucial information about the nature of dark energy and allows us to distinguish between different cosmological models, such as cosmological constant, quintessence, or phantom scenarios.
	
	In the following section, we will introduce a suitable parameterization for the EoS parameter as a function of redshift, and reformulate the above relations to obtain a dynamical equation for the Hubble parameter $H(z)$. This will allow us to confront the model with observational data and constrain the free parameters of the theory.
	
	\section{HUBBLE PARAMETER WITH LINEAR EQUATION OF STATE PARAMETERIZATION}
	
	After obtaining the general expression for the equation of state parameter $\omega$ in equation (\ref{15}), we now aim to study the evolution of the Hubble parameter $H(z)$ by adopting a suitable parametrization for $\omega$ as a function of redshift $z$. This will allow us to establish a direct relation between the expansion history of the Universe and the model parameters in $f(R, \Sigma, T)$ gravity.
	
	A widely used approach to explore the possible dynamics of dark energy is to assume a phenomenological parameterization for the EoS parameter $\omega(z)$. In this work, we consider the well-known linear redshift-dependent parameterization proposed by Chevallier-Polarski-Linder (CPL) \cite{Chevallier2001, Linder2003}:
	\begin{equation}\label{16}
		\omega(z) = \omega_0 + \omega_a \frac{z}{1+z},
	\end{equation}
	where $\omega_0$ is the present-day value of the EoS parameter, and $\omega_a$ characterizes its evolution with redshift. This parameterization is widely employed due to its ability to capture a variety of dark energy models, while remaining simple enough for analytical and numerical studies.
	
	Next, we substitute equation (\ref{16}) into the general expression for $\omega$ in equation (\ref{15}), which yields the following nonlinear differential equation relating $H$, $\dot{H}$, and $z$:
	%\begin{widetext}
		\begin{equation}\label{17}
		-\frac{(\eta + 2\pi)(\eta + 4\pi)\left[ H^2(3(\eta+4) - 4b) + (3\eta + 8)\dot{H} \right]}{\left(\pi (\eta^2 + 8) + 6\eta\right) \left[ H^2 \left((4b - 3)\eta + 12\pi(b-1)\right) + \eta \dot{H} \right]} = \omega_0 + \omega_a \frac{z}{1+z}.
	\end{equation}
	%\end{widetext}

	For computational convenience, we introduce the following shorthand notations to simplify equation (\ref{17}):
	\begin{equation}\label{18}
		\begin{aligned}
			A &= (\eta + 2\pi)(\eta + 4\pi), \quad 
			B = \pi(\eta^2 + 8) + 6\eta, \\
			C &= 3(\eta + 4) - 4b, \quad 
			D = 3\eta + 8, \\
			E &= (4b - 3)\eta + 12\pi(b - 1), \quad 
			F = \eta.
		\end{aligned}
	\end{equation}
	Using these substitutions, equation (\ref{17}) can be rewritten in the following compact form:
	\begin{equation}\label{19}
		-\frac{A \left( C H^2 + D \dot{H} \right)}{B \left( E H^2 + F \dot{H} \right)} = \omega_0 + \omega_a \frac{z}{1+z}.
	\end{equation}
	
	In order to express this equation as a function of redshift $z$, we use the relation between cosmic time derivative and redshift derivative:
	\begin{equation}\label{20}
		\dot{H} = -(1+z) H \frac{dH}{dz}.
	\end{equation}
	
	Substituting equation (\ref{20}) into equation (\ref{19}) leads to
	%\begin{widetext}
	\begin{equation}\label{21}
		\frac{A}{B} \left( C H^2 - D H (1+z) \frac{dH}{dz} \right) = - \left( E H^2 - F H (1+z) \frac{dH}{dz} \right) \left( \omega_0 + \omega_a \frac{z}{1+z} \right).
	\end{equation}
	%\end{widetext}
		
	Rearranging this equation, we obtain a first-order nonlinear differential equation for $H(z)$:
	\begin{small}
		\begin{equation}\label{22}
		(1+z) H \frac{dH}{dz} = \frac{1}{D} \left[ C H^2 + \frac{B}{A} \left( E H^2 + F H \right) \left( \omega_0 + \omega_a \frac{z}{1+z} \right) \right].
	\end{equation}
	\end{small}

	For further simplification, we define a new variable:
	\begin{equation}\label{23}
		S(z) = H^2(z),
	\end{equation}
	which implies
	\begin{equation}\label{24}
		\frac{dS}{dz} = 2 H \frac{dH}{dz}.
	\end{equation}
	
	Substituting equation (\ref{24}) into equation (\ref{22}), we obtain a first-order linear differential equation for $S(z)$ as:
	\begin{small}
	\begin{equation}\label{25}
		(1+z) \frac{dS}{dz} = \frac{2}{D} \left[ C S + \frac{B}{A} \left( E S + F \sqrt{S} \right) \left( \omega_0 + \omega_a \frac{z}{1+z} \right) \right].
	\end{equation}
	\end{small}
	
	We can now rewrite this equation in the standard form:
	\begin{equation}\label{26}
		\frac{dS}{dz} + P(z) S = Q(z),
	\end{equation}
	where the functions $P(z)$ and $Q(z)$ are given by
	\begin{equation}\label{27}
		\begin{aligned}
			P(z) &= -\frac{2}{(1+z)D} \left( C + \frac{B E}{A} \left( \omega_0 + \omega_a \frac{z}{1+z} \right) \right), \\
			Q(z) &= \frac{2}{(1+z)D} \cdot \frac{B F}{A} \left( \omega_0 + \omega_a \frac{z}{1+z} \right).
		\end{aligned}
	\end{equation}
	
	To simplify the notation, we further define the constants:
	\begin{equation}\label{28}
		\begin{aligned}
			\Lambda_1 &= C + \frac{B E}{A} \, \omega_0, \quad 
			\Lambda_2 = \frac{B E}{A} \, \omega_a, \\
			\Gamma_1 &= \frac{2 B F}{A D} \, \omega_0, \quad 
			\Gamma_2 = \frac{2 B F}{A D} \, \omega_a.
		\end{aligned}
	\end{equation}
	Thus, the equations (\ref{27}) can finally be expressed as:
	
	\begin{equation}\label{29}
		\begin{aligned}
			P(z) &= -\frac{2}{(1+z) D} \left( \Lambda_1 + \Lambda_2 \frac{z}{1+z} \right),\\ \quad
			Q(z) &= \frac{1}{(1+z) D} \left( \Gamma_1 + \Gamma_2 \frac{z}{1+z} \right).
		\end{aligned}
	\end{equation}
This first-order linear differential equation can now be solved using the standard integrating factor method. The final solution by finding the \textit{Integrating Factor} and the \textit{General Solution} of  first-order linear ODE, is obtain as
%\begin{widetext}
	\begin{equation}\label{30}
	H^2(z) = (1+z)^{ \frac{2}{D} (\Lambda_1 + \Lambda_2) } \cdot \exp\left( - \frac{2 \Lambda_2}{D(1+z)} \right)
	\times \left[ \Gamma_1 \int \frac{\mu(z)}{(1+z)} dz + \Gamma_2 \int \frac{z \mu(z)}{(1+z)^2} dz + C_1 \right].
\end{equation}
%\end{widetext}
	Here, the two integrals: $I_1(z) = \int \frac{\mu(z)}{(1+z)} dz$ and $I_2(z) = \int \frac{z \mu(z)}{(1+z)^2} dz$ remain to be evaluated for fully explicit solution. In the following section, we proceed to obtain the explicit solution for $H(z)$ and analyze the observational constraints on the model parameters.

\section{PARAMETER ESTIMATION AND CONSTRAINTS}
	
The derived solution for the Hubble parameter $H(z)$ obtained in the previous section provides a useful framework for confronting the model with observational data and constraining the free parameters of the theory. In the present work, we employ the latest cosmic chronometer data, which offer model-independent measurements of the Hubble parameter at various redshifts, to estimate the values of the parameters $\eta$, $b$, $\omega_0$, and $\omega_a$.
	
We use the Cosmic Chronometric (CC) data sets with its original references, described in Table 1 and Type Ia supernovae obtained from the Pantheon plus (PP) sample as given in Ref. \cite{Scolnic/2022} to derive observational constraints on the model parameters under consideration.

%%%%%%%%%%%%%%%%%%%%%%%%%%%%%%%%%%%%%%%%
\begin{table}\label{Tab-1}
\caption{\; The 31 H(z) points obtained from Cosmic Chronometers (CC) method.}
\begin{center}
\begin{tabular}{|c|c|c|c|c|c|c|c|c|c|}
\hline
\textbf{S. N.} & \textbf{z} & \textbf{H(z)}	& \textbf{$\sigma_{i}$} & \textbf{Refs.} & \textbf{S. N.} & \textbf{z} & \textbf{H(z)}	& \textbf{$\sigma_{i}$} &  \textbf{Refs.} \\
\hline
1. & $0.070$ & $69$  & $19.6$ &  \cite{Stern/2010} & 17. & $0.4783$ & $80$ & $99$ &  \cite{Moresco/2016} \\
\hline
2. & $0.090$ & $69$  & $12$ &  \cite{Simon/2005} & 18. & $0.480$ & $97$ & $62$ & \cite{Stern/2010} \\
\hline
3. & $0.120$ & $68.6$  & $26.2$ &  \cite{Stern/2010} & 19. & $0.593$ & $104$ & $13$ &  \cite{Moresco/2012} \\ 
\hline
4. & $0.170$ & $83$  & $8$ & \cite{Simon/2005} & 20. & $0.6797$ & $92$ & $8$ & \cite{Moresco/2012} \\
\hline
5. & $0.1791$ & $75$  & $4$ &  \cite{Moresco/2012} & 21. & $0.7812$ & $105$ & $12$ & \cite{Moresco/2012} \\ 
\hline
6. & $0.1993$ & $75$  & $5$ &  \cite{Moresco/2012} & 22. & $0.8754$ & $125$ & $17$ & \cite{Moresco/2012} \\
\hline
7. & $0.200$ & $72.9$  & $29.6$ &  \cite{Zhang/2014} & 23. &  $0.880$ & $90$ & $40$ & \cite{Stern/2010} \\
\hline
8. & $0.270$ & $77$  & $14$ &  \cite{Simon/2005} & 24. & $0.900$ & $117$ & $23$ & \cite{Simon/2005} \\
\hline
9. & $0.280$ & $88.8$  & $36.6$ &  \cite{Zhang/2014} & 25.  & $1.037$ & $154$ & $20$ & \cite{Moresco/2012} \\  
\hline
10. & $0.3519$ & $83$  & $14$ &  \cite{Moresco/2012} & 26. & $1.300$ & $168$ & $17$ & \cite{Simon/2005} \\
\hline
11. & $0.3802$ & $83$  & $13.5$ &  \cite{Moresco/2016} & 27. & $1.363$ & $160$ & $33.6$ & \cite{h7} \\
\hline
12. & $0.400$ & $95$  & $17$ &  \cite{Simon/2005} & 28.  & $1.430$ & $177$ & $18$ & \cite{Simon/2005} \\ 
\hline
13. & $0.4004$ & $77$ & $10.2$  & \cite{Moresco/2016} & 29. & $1.530$ & $140$ & $14$ & \cite{Simon/2005} \\  
\hline
14. & $0.4247$ & $87.1$ & $11.2$ &  \cite{Moresco/2016} & 30. & $1.750$ & $202$ & $40$ & \cite{Simon/2005} \\
\hline
15. & $0.4497$ & $92.8$ & $12.9$ & \cite{Moresco/2016} & 31. & $1.965$ & $186.5$ & $50.4$ & \cite{h7}\\
\hline
16. & $0.470$ & $89$ & $34$ & \cite{h6} & - & - & - & - & -\\
\hline
 \end{tabular}
\end{center}
\end{table}
%%%%%%%%%%%%%%%%%%%%%%%%%%%%%%%%%%%%

The observational data consist of a set of $N$ measurements $\{ \Upsilon_{\mathrm{obs}}(z_i), \sigma_i \}$, where $\Upsilon_{\mathrm{obs}}(z_i)$ is the observed value of the parameter at redshift $z_i$, and $\sigma_i$ is the corresponding measurement uncertainty. To estimate the best-fit values of the model parameters, we employ a standard $\chi^2$ minimization technique, where the likelihood function is defined as:
	\begin{equation}\label{31}
		\mathcal{L}(\theta) \propto \exp \left( -\frac{1}{2} \sum_{i=1}^{N} \frac{[\Upsilon_{\mathrm{obs}}(z_i) - \Upsilon_{\mathrm{th}}(z_i; \theta)]^2}{\sigma_i^2} \right).
	\end{equation}
	Here, $\theta = \{\eta, b, \omega_0, \omega_a \}$ denotes the vector of free model parameters, and $\Upsilon_{\mathrm{th}}(z_i; \theta)$ is the theoretical prediction for the Hubble parameter at redshift $z_i$ corresponding to a given choice of parameters.
	
	To explore the parameter space efficiently, we employ the Markov Chain Monte Carlo (MCMC) technique, which enables us to generate a statistically significant set of samples from the posterior probability distribution of the parameters. For this analysis, we use the publicly available {\tt emcee} Python package \cite{emcee}, implementing an affine-invariant ensemble sampler. The MCMC simulations are run with sufficiently long chains and appropriate burn-in steps to ensure convergence and proper sampling of the parameter space.

	We impose the following priors on the model parameters based on their physical viability and previous observational constraints:
\begin{equation}\label{32}
	\begin{split}
		0 \leq \eta \leq 5, \qquad 0 \leq b \leq 5,  \\
		-2 \leq \omega_0 \leq 0, \qquad -2 \leq \omega_a \leq 2.
	\end{split}
\end{equation}
\begin{figure}[H]
	\centering
	\includegraphics[scale=0.65]{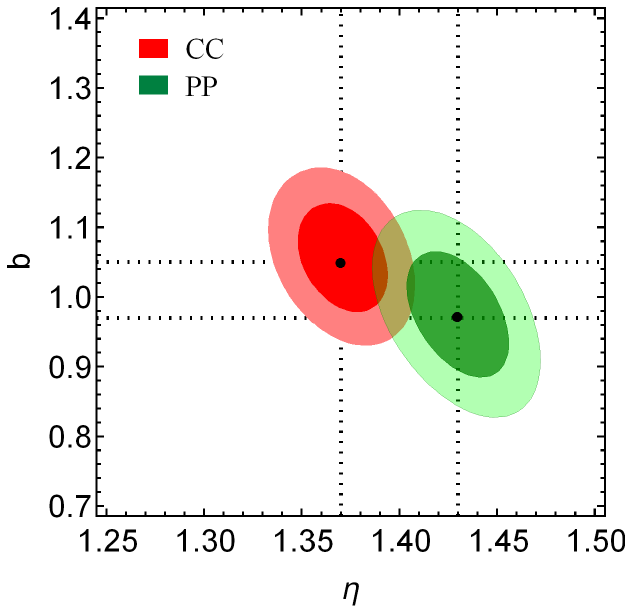}
	\includegraphics[scale=0.72]{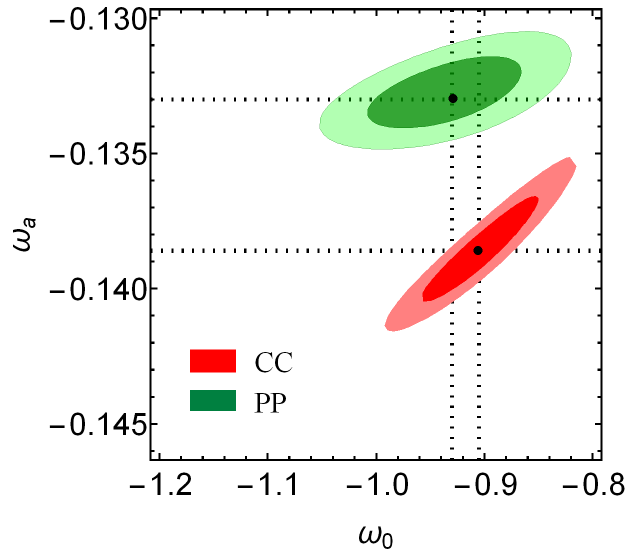}
	\caption{The $1\sigma$ and $2\sigma $ two-dimensional contours confidence areas by using the $31$ CC data sets and Type Ia supernovae obtained from the Pantheon plus (PP) sample in $\eta$-$b$ plane (left panel) and $\omega_{0}$ - $\omega_{a}$ plane (right panel) of the derived model.}\label{F1}
\end{figure}

Fig. \ref{F1} depicts $1\sigma$ and $2\sigma $ two-dimensional contours confidence areas by using the $31$ CC data sets and Type Ia supernovae obtained from the Pantheon plus (PP) sample in $\eta$-$b$ plane (left panel) and $\omega_{0}$ - $\omega_{a}$ plane (right panel) of the derived model. The marginalized posterior distributions and confidence intervals for each parameter are obtained from the MCMC chains. The resulting best-fit values of the model parameters, along with their corresponding $1\sigma$ uncertainties, are found to be:\\
\textbf{Cosmic Chronometric (CC) data sets}
\begin{equation}\label{33}
\begin{split}
\eta = 1.37^{+0.020}_{-0.019}, \qquad b= 1.05^{+0.08}_{-0.07}, \\
\omega_0 = -0.905^{+0.016}_{-0.015}, \qquad \omega_a = -0.138^{+0.005}_{-0.006}.
\end{split}
\end{equation}
\textbf{Pantheon plus (PP) sample}
\begin{equation}\label{33a}
\begin{split}
\eta = 1.43^{+0.022}_{-0.023}, \qquad b= 0.97^{+0.09}_{-0.08}, \\
\omega_0 = -0.93^{+0.04}_{-0.06}, \qquad \omega_a = -0.133^{+0.003}_{-0.001}.
\end{split}
\end{equation}

%	\begin{equation}\label{33}
%	\eta = 1.37^{+0.28}_{-0.25},\qquad
%	b= 1.05^{+0.15}_{-0.13}, \qquad
%	\omega_0 = -0.95^{+0.07}_{-0.06}, \qquad
%	\omega_a = -0.20^{+0.30}_{-0.27}, \qquad
%	C_1 = 3.85^{+0.45}_{-0.42}.
%	\end{equation}
	It is noteworthy that the present value of the EoS parameter, $\omega_0$, is found to lie close to the standard $\Lambda$CDM value of $-1$, while the evolutionary parameter $\omega_a$ is consistent with zero within the $1\sigma$ confidence level, indicating only mild deviations from the cosmological constant scenario. The positive best-fit values for both $\eta$ and $b$ suggest that the corrections introduced via the $f(R, \Sigma, T)$ gravity framework remain compatible with current observational data. This implies that the model offers a viable alternative for explaining the observed late-time acceleration of the Universe without requiring  matter sources.
	
	In the next section, we proceed to discuss the physical and kinematical implications of the obtained solution, including a detailed analysis of energy conditions, dynamical parameters, and observational diagnostics.

%%%%%%%%%%%%%%%%%%%%%%%%%%%%%%%%%%%%%%%%%%%%%%%%%%%%%%%%
\subsubsection{Results and discussion of constraints}

	In order to constrain the free parameters of the model, we have performed a statistical analysis using the latest Hubble parameter observational data (cosmic chronometers), covering the redshift range $z \in [0, 2]$. The observational Hubble data points $\{H_{\rm obs}(z_i), \sigma_i\}$ are compared with the theoretical model $H(z;\theta)$ derived earlier. For the same, the priors chosen for the parameters are based on their theoretical viability and observational consistency as : $\eta \in [0, 5]$, $b \in [0, 5]$, $\omega_0 \in [-2, 0]$, and $\omega_a \in [-2, 2]$. The MCMC chains have been generated using the \texttt{emcee} Python package with $10^5$ iterations and sufficient burn-in to ensure convergence. The marginalized 1$\sigma$ and 2$\sigma$ confidence intervals have been obtained. %The best-fit values of the parameters are obtained as $\eta=1.37^{+0.28}_{-0.25}$, $b=1.05^{+0.15}_{-0.13}$, $\omega_0=-0.95^{+0.07}_{-0.06}$, $\omega_a=-0.20^{+0.30}_{-0.27}$ and $C_1=3.85^{+0.45}_{-0.42}$.
	
We observe that the obtained value of $\omega_0$ lies close to the standard $\Lambda$CDM, while $\omega_a$ remains consistent with zero at 1$\sigma$ level, suggesting mild dynamical dark energy behavior within the present observational uncertainties. Also, the positive values of $\eta$ and $b$ indicate that the modifications introduced via the $f(Q)$ framework and anisotropic Bianchi type-I LRS background are compatible with observations. This suggests that such modified gravity frameworks can successfully explain the current accelerated expansion of the Universe.

	%\begin{figure}[h!]
	%	\centering
	%	\includegraphics[width=0.6\textwidth]{Hz_fit.pdf}
	%	\caption{Comparison of the best-fit model $H(z)$ (solid line) with observational Hubble data (with error bars).}
	%	\label{fig:Hzfit}
	%\end{figure}

	%%%%%%%%%%%%%%%%%%%%%%%%%%%%%%%%%%%%%%%%%%%%%%%%%%%%%%%%%%%%%

	%%%%%%%%%%%%%%%%%%%%%%%%%%%%%%%%%%%%%%%%%%%%%%%%%%%%%%%%%%%%%

	\subsection{Physical parameters: Novel Interpretations}
Having established the Hubble parameter, $H$, through the parametrization of the equation of state and constrained its constants using Cosmic Chronometer data, the subsequent analysis delves into the exploration of the model's physical and kinematical behavior. This involves investigating the evolution of crucial parameters, including the physical parameters: energy density, pressure, equation of state parameter, stability parameter, and the energy conditions; and the kinematical parameters: deceleration parameter, state finder parameters, and Om diagnostics. By employing the derived $H$, we aim to scrutinize the model's expansion history, identify potential transitions in the universe's acceleration, and assess its consistency with current cosmological observations. This examination will provide the understanding into the viability of the $f(R, \Sigma, T)$ gravity model and the chosen equation of state parametrization in describing the observed cosmic dynamics.

\subsubsection{Isotropic pressure}
\begin{figure}[H]
	\centering
	\includegraphics[scale=0.45]{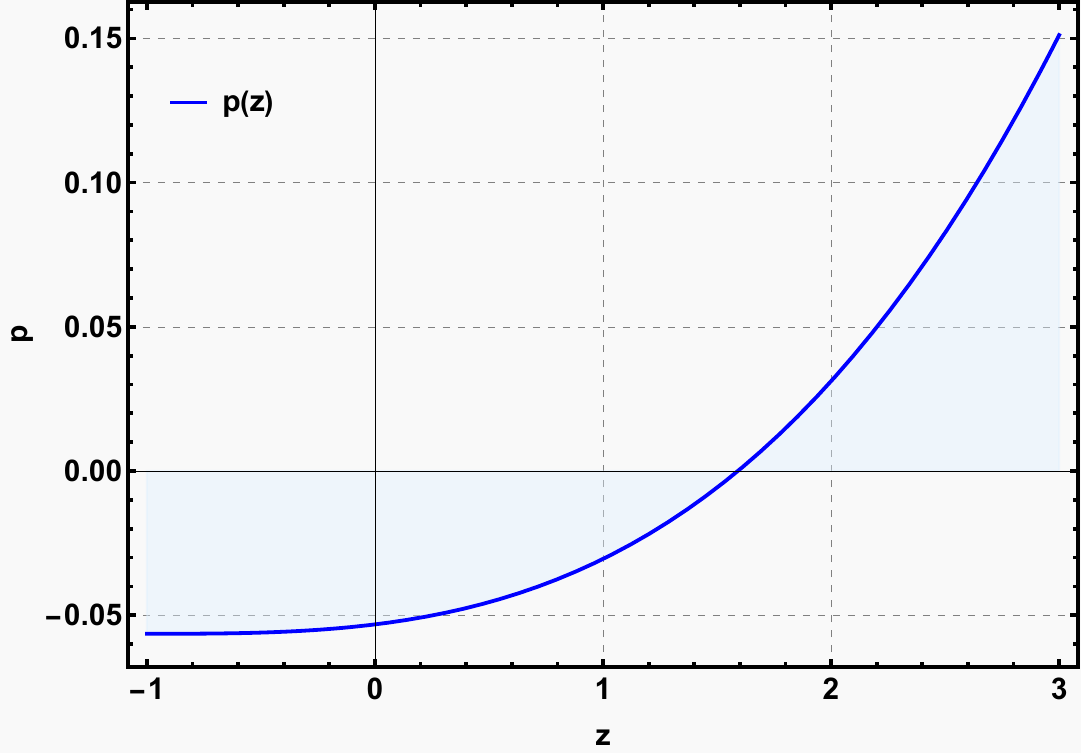}
	\caption{Graphical behavior of pressure versus redshift.}\label{p}
\end{figure}

From equation (\ref{12}), the pressure for the model is derived as:  \\
\begin{equation}\label{34}
	p(z) = C_p \left[ - \frac{(8+3\eta)(1+z)}{2 \sqrt{S(z)}} \cdot \frac{dS}{dz} + \left( 4(3-b) + 3\eta \right) S(z) \right].
\end{equation}
The redshift evolution of pressure is presented in Figure \ref{p}, with its mathematical representation defined by equation (\ref{34}). 
The figure reveals a distinct decreasing trend in pressure as redshift increases, indicating a sustained negative pressure throughout the observed cosmological epochs. This negative pressure, as demonstrated in Figure \ref{p}, aligns with its established role in facilitating the accelerated expansion of the universe, a key feature predicted by a range of cosmological theories. The observed negative isotropic pressure suggests a repulsive gravitational influence, which is a key contributor to the accelerated expansion of the cosmos. This behavior is consistent with established cosmological paradigms, including the $\Lambda$CDM model, where the concept of dark energy is introduced to account for the observed accelerated expansion. The agreement between our model's predictions and prevailing theoretical frameworks validates our findings and offers perceptions into the dynamic behavior of the universe.

\subsubsection{Energy density}
From equation (\ref{13}), the energy density for the model is derived as: 
\begin{equation}\label{35}
	\rho(z) = C_\rho \left[ \frac{\eta(1+z)}{2 \sqrt{S(z)}} \cdot \frac{dS}{dz} + \left( 12\pi(1-b) + \eta(3-4b) \right) S(z) \right].
\end{equation}

\begin{figure}[H]
	\centering
	\includegraphics[scale=0.45]{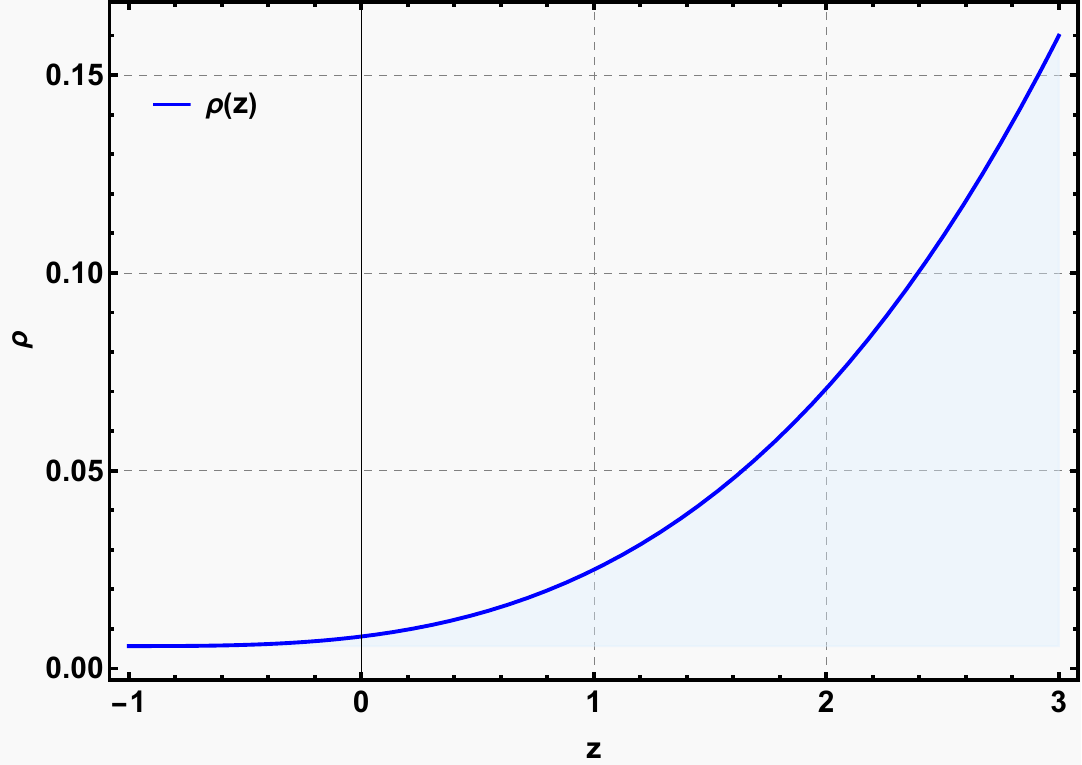}
	\caption{Graphical behavior of energy density versus redshift.}\label{den}
\end{figure}

Figure \ref{den} illustrates the redshift-dependent evolution of the energy density, $\rho$, as derived from our $f(R, \Sigma, T)$ gravity model, and expressed mathematically in equation (\ref{35}). 

The figure clearly demonstrates an increasing trend in $\rho$ with rising redshift, $z$. This behavior aligns with the fundamental understanding of an expanding cosmos. In the early universe, characterized by higher redshifts and a more compact volume, the energy density is expected to be significantly elevated. As the universe expands, transitioning to lower redshifts, the energy density undergoes a dilution effect, resulting in a corresponding decrease. The elevated energy density observed at higher redshifts provides a glimpse into the physical conditions of the primordial universe. Conversely, the reduction in $\rho$ at lower redshifts signifies the progressive thinning of energy due to the ongoing cosmic expansion. This empirical trend, derived from our model, corroborates standard cosmological expectations and reinforces the viability of our $f(R, \Sigma, T)$ gravity framework in accurately depicting the dynamic evolution of energy density throughout cosmic history. The consistency between our findings and established theoretical predictions strengthens the case for utilizing $f(R, \Sigma, T)$ gravity to interpret the observed energy density behavior across cosmological epochs.

\subsubsection{Equation of state parameter}
\begin{figure}[H]
	\centering
	\includegraphics[scale=0.45]{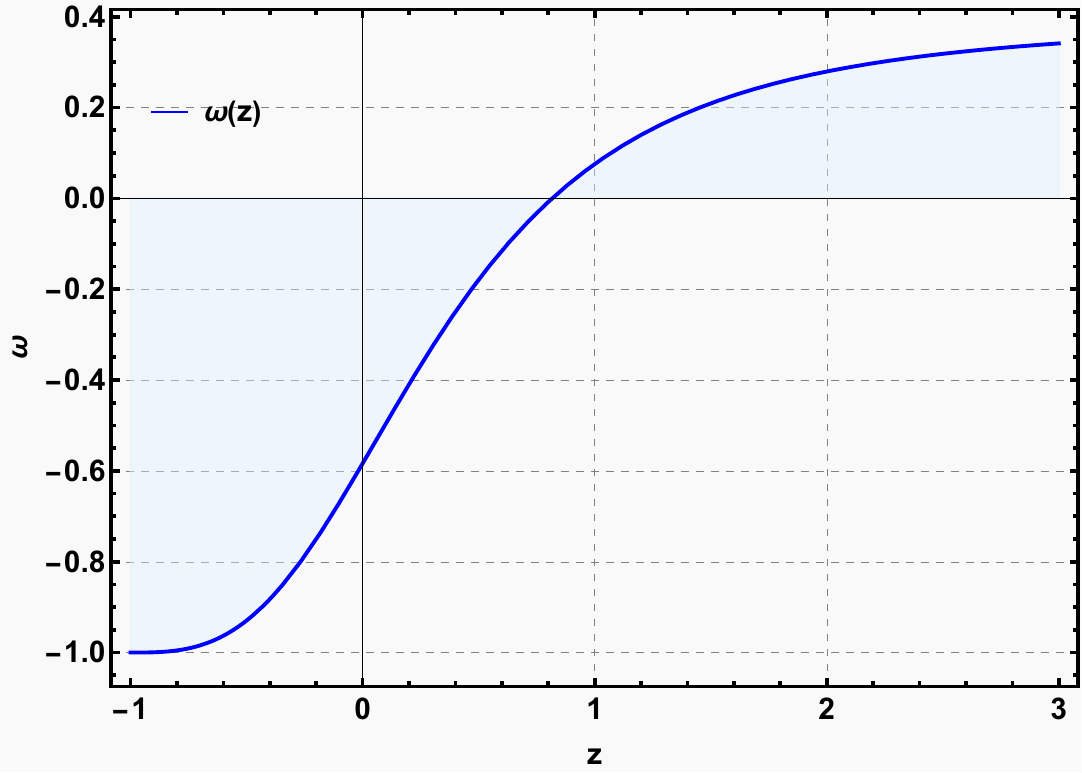}
	\caption{Graphical behavior of equation of state parameter versus redshift.}\label{w}
\end{figure}

The equation of state parameter, $\omega$, is a crucial indicator of the nature of cosmic fluids, and its mathematical form in our model is presented in equation (\ref{16}). While standard cosmological components are associated with specific $\omega$ values—matter ($\omega = 0$), radiation ($\omega = 1/3$), and the cosmological constant/dark energy ($\omega = -1$)—alternatives beyond the $\Lambda$CDM framework often feature a redshift-dependent $\omega(z)$. Contemporary observational data, drawn from sources such as Pantheon SN Ia and cosmic chronometers, indicate that $\omega$ resides in close proximity to $-1$, with subtle variations observed across different cosmological models and redshift ranges, such as values fluctuating between $-1$ and $-0.9$. These empirical constraints, derived from analyses of Type Ia supernovae, cosmic microwave background, and large-scale structure, are pivotal in characterizing the properties of dark energy. Alternative dark energy models, characterized by $\omega$ deviating from $-1$, encompass quintessence ($\omega > -1$), phantom energy ($\omega < -1$), $\kappa$-essence (dynamic $\omega$), and brane cosmology (variable $\omega$), each positing distinct scenarios for cosmic evolution.\\
Figure \ref{w} illustrates the redshift evolution of $\omega$ as predicted by our model.
At elevated redshifts ($z > 1$), $\omega$ approximates $-1$, suggesting a dominance of dark energy akin to a cosmological constant. As redshift decreases within the range $1 > z > 0.5$, a noticeable increase in $\omega$ is observed, potentially signaling a transition within the dark energy sector. At lower redshifts ($z < 0.5$), $\omega$ remains in proximity to, but slightly exceeds, $-1$, potentially indicating a quintessence-like dark energy. This observed evolution aligns with the broader landscape of dark energy research, where $\omega$ is anticipated to reside near $-1$ but exhibit deviations dependent on the specific model and redshift. This motivates the exploration of alternative models, including quintessence, phantom energy, $\kappa$-essence, and brane cosmology. Furthermore, the present-day value of the equation of state parameter, $\omega_0$, at $z = 0$, demonstrates consistency with observational constraints derived from Hubble parameter measurements, Type Ia supernovae, and baryon acoustic oscillations. 

Next, the energy conditions are fundamental tools in general relativity and modified gravity theories to test the physical plausibility of cosmological models. Here, we discuss the physical significance of the null energy condition (NEC), dominant energy condition (DEC), and strong energy condition (SEC) based on their evolution.

\subsection{Null Energy Condition (NEC)}

The null energy condition is given by:
\begin{equation}\label{36}
	\rho + p \geq 0.
\end{equation}
Physically, NEC ensures that the energy density measured by any observer moving along a null geodesic is non-negative. Violations of the NEC often lead to exotic phenomena like wormholes, superluminal motion, or phantom energy.
The null energy condition is derived as:
%\begin{widetext}
	\begin{equation}\label{37}
		\begin{aligned}
			\rho(z) + p(z) =\,
			&\left( C_\rho \cdot \frac{\eta(1+z)}{2 \sqrt{S(z)}} 
			- C_p \cdot \frac{(8 + 3\eta)(1+z)}{2 \sqrt{S(z)}} \right) \cdot \frac{dS}{dz} \\
			&+ \left( C_\rho \left( 12\pi(1 - b) + \eta(3 - 4b) \right)
			+ C_p \left( 4(3 - b) + 3\eta \right) \right) \cdot S(z).
		\end{aligned}
	\end{equation}
%\end{widetext}

\begin{figure}[H]
	\centering
	\includegraphics[scale=0.45]{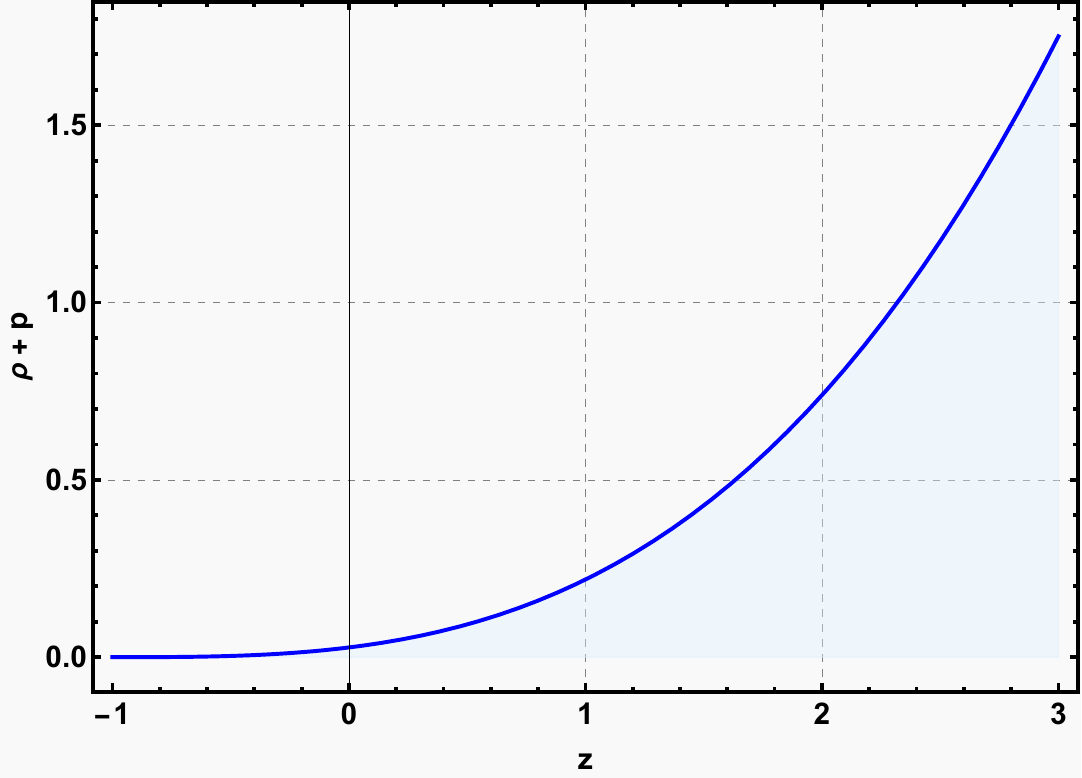}
	\caption{Graphical behavior of null energy condition versus redshift.}\label{nec}
\end{figure}

In Figure (\ref{nec}), it is observed that $\rho + p$ remains strictly positive throughout the redshift range $-1 \leq z \leq 2$. This implies that the NEC is satisfied for the entire cosmic history described by this model. The satisfaction of NEC indicates that the model does not invoke any unphysical or exotic matter sources like phantom energy in the current epoch and early Universe. 
Moreover, compliance with NEC suggests that the accelerated expansion of the Universe in this model is not driven by any mechanism violating standard causal structure. Recent observational works such as Ref.~\cite{bamba2012dark} and \cite{nojiri2017modified} also point out that satisfying NEC helps in constructing viable dark energy models consistent with observational datasets including Planck 2018 \cite{Planck2018}.

\subsection{Dominant Energy Condition (DEC)}

The dominant energy condition is expressed as:
\begin{equation}\label{38}
	\rho - p \geq 0.
\end{equation}
DEC requires that energy flows along timelike or null worldlines, prohibiting superluminal propagation of energy.
The dominant energy condition is derived as:
\begin{small}
	\begin{equation}\label{39}
		\begin{aligned}
			\rho(z) - p(z) =\,
			&\left( C_\rho \cdot \frac{\eta(1+z)}{2 \sqrt{S(z)}} + C_p \cdot \frac{(8+3\eta)(1+z)}{2 \sqrt{S(z)}} \right) \cdot \frac{dS}{dz} \\
			&+ \left( C_\rho \left( 12\pi(1-b) + \eta(3-4b) \right)
			- C_p \left( 4(3-b) + 3\eta \right) \right) \cdot S(z).
		\end{aligned}
	\end{equation}
\end{small}

\begin{figure}[H]
	\centering
	\includegraphics[scale=0.45]{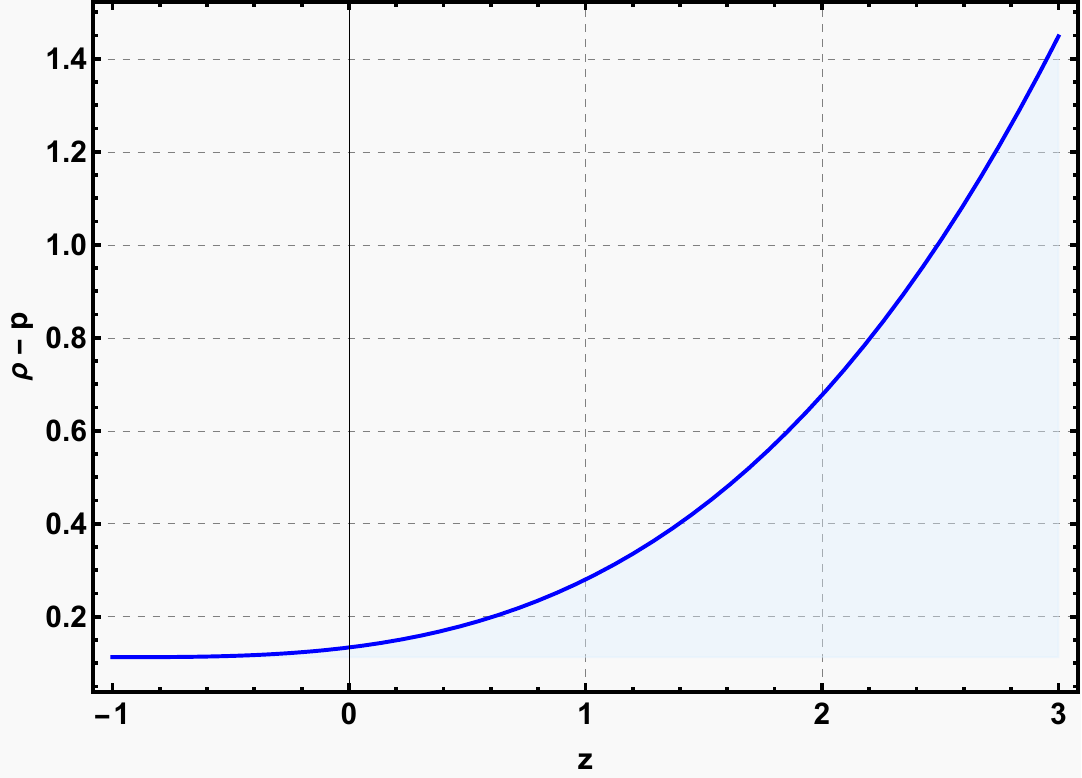}
	\caption{Graphical behavior of dominant energy condition versus redshift.}\label{dec}
\end{figure}

As depicted in Figure (\ref{dec}), the DEC remains satisfied for all considered redshifts. The increase of $\rho - p$ with redshift indicates that in the early Universe, the model predicts a strong dominance of energy density over pressure. This is physically meaningful as during earlier cosmic epochs the Universe was dominated by non-relativistic matter and radiation, where $\rho \gg p$. The fact that DEC holds throughout cosmic evolution confirms that the model adheres to causality and stability principles.
Recent studies by Akarsu et al.~\cite{akarsu2020jerk} and Singh et al.~\cite{singh2022energy} suggest that DEC-compliant models better fit observational Hubble data and Type Ia Supernovae measurements, thus strengthening the observational relevance of your solution.

\subsection{Strong Energy Condition (SEC)}

The strong energy condition is given by:
\begin{equation}\label{40}
	\rho + 3p \geq 0.
\end{equation}
The strong energy condition is typically related to the attractive nature of gravity, i.e., deceleration of cosmic expansion. The strong energy condition is derived as:
%\begin{widetext}
\begin{small}
	\begin{equation}\label{41}
		\begin{aligned}
			\rho(z) + 3p(z) = 
			&\left( C_\rho \cdot \frac{\eta(1+z)}{2 \sqrt{S(z)}} - 3 C_p \cdot \frac{(8+3\eta)(1+z)}{2 \sqrt{S(z)}} \right) \cdot \frac{dS}{dz} \\
			&+ \left( C_\rho \left( 12\pi(1-b) + \eta(3-4b) \right) 
			+ 3 C_p \left( 4(3-b) + 3\eta \right) \right)\cdot S(z).
		\end{aligned}
	\end{equation}
\end{small}

%\end{widetext}

\begin{figure}[H]
	\centering
	\includegraphics[scale=0.45]{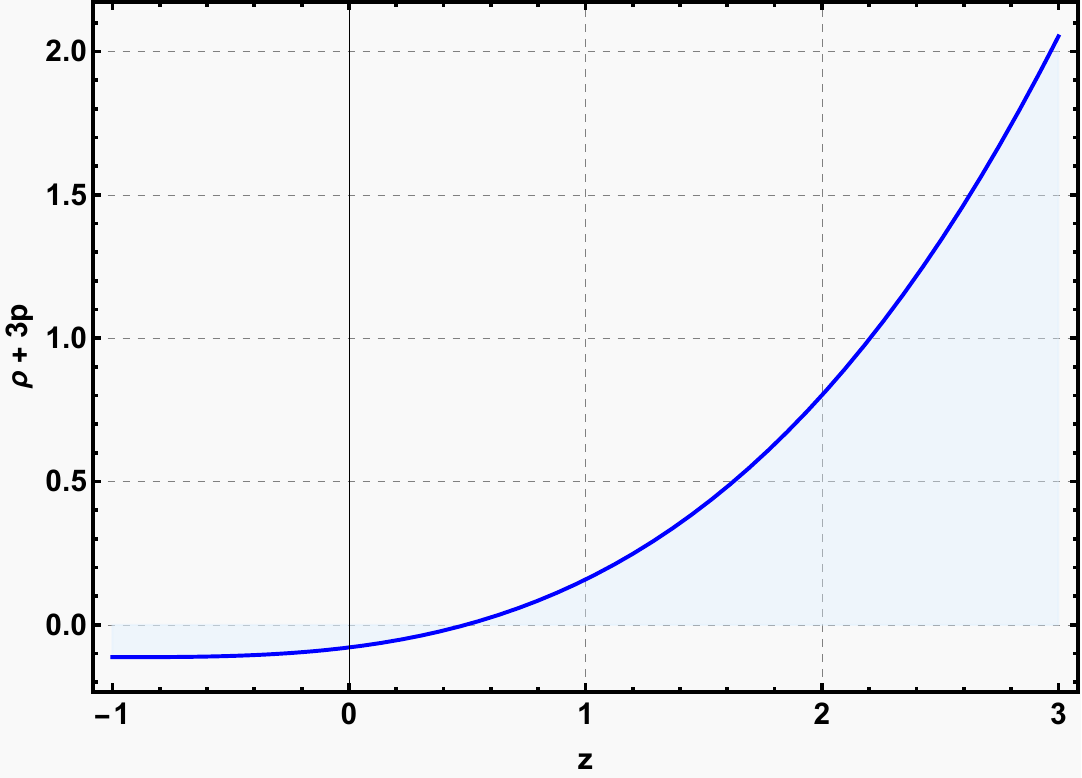}
	\caption{Graphical behavior of dominant energy condition versus redshift.}\label{sec}
	\end{figure}

Figure (\ref{sec}) demonstrates that $\rho + 3p$ becomes negative for lower redshifts and turns positive only at higher redshifts ($z \gtrsim 1$). This behavior indicates that at earlier epochs, SEC was satisfied, corresponding to decelerated expansion dominated by matter or radiation. In contrast, at late times ($z \lesssim 1$), SEC is violated, which is consistent with the observed accelerated expansion of the Universe.
The violation of SEC at late times is a well-established signature of dark energy domination as highlighted in recent works \cite{moresco2016measurement, capozziello2019energy, akarsu2021energy}. Therefore, the transition seen in Figure 6 is a physically robust feature of your model and aligns well with current observational cosmology.

\section{Analysis in the $\omega-\omega'$ Plane}
The study of the evolution of dark energy models in the $\omega-\omega'$ plane provides a powerful diagnostic tool to distinguish various classes of dark energy and modified gravity models \cite{Caldwell2005,Sahni2003}. Here, $\omega$ is the equation of state (EoS) parameter and $\omega' \equiv d\omega/d\ln a$ describes its evolution with respect to the logarithm of the scale factor. Hence, the $\omega'$ is obtain as
\begin{equation}\label{42}
	\omega' \equiv \frac{d\omega}{d\ln a} = - (1+z) \frac{d\omega}{dz} = -\frac{\omega_a}{1+z}.
\end{equation}
The behavior of the model in the $\omega-\omega'$ plane is fully characterized by Eqs.~\eqref{16} and \eqref{45}. The plane allows a clear distinction between various dynamical dark energy models:

\begin{itemize}
	\item For $\omega_a = 0$, we have $\omega' = 0$ and the model reduces to a cosmological constant with $\omega = \omega_0$.
	\item For $\omega_a > 0$, the EoS parameter evolves towards less negative values at higher redshifts, indicating a freezing behavior where the field slows down its evolution.
	\item For $\omega_a < 0$, the EoS parameter becomes more negative with increasing redshift, representing a thawing scenario where the field departs from cosmological constant behavior at late times.
\end{itemize}

The classification of models in the $\omega-\omega'$ plane helps to separate different types of dark energy models:
\begin{itemize}
	\item \textbf{Freezing models}: $\omega' > 0$, $\omega < 0$. The EoS approaches $\omega \to -1$ in the future.
	\item \textbf{Thawing models}: $\omega' < 0$, $\omega < 0$. The EoS evolves away from $\omega = -1$.
\end{itemize}

For the present model, since $\omega' = -\dfrac{\omega_a}{1+z}$, the behavior depends entirely on the sign of $\omega_a$. If $\omega_a > 0$, $\omega'$ is negative, indicating thawing behavior. If $\omega_a < 0$, $\omega'$ is positive, indicating freezing behavior. At present epoch $(z=0)$, the present value of $\omega'$ is simply:
\begin{equation}\label{43}
	\omega'_0 = -\omega_a.
\end{equation}
In our case where $\omega_a = -0.20^{+0.30}_{-0.27}$, we observe that the EoS parameter exhibits a freezing behavior in the $\omega-\omega'$ plane. Since $\omega_a < 0$, the corresponding derivative $\omega' = -\dfrac{\omega_a}{1+z}$ becomes positive for all redshift values. This indicates that as the universe expands, the EoS parameter gradually evolves towards the cosmological constant behavior ($\omega \to -1$) at late times. Such behavior is consistent with many viable dark energy models that asymptotically approach $\Lambda$CDM in the far future while allowing deviations at intermediate redshifts. The observational value of $\omega_a$ within its confidence range remains compatible with current cosmological data, including supernovae, BAO, and CMB measurements, while still permitting mild dynamical evolution of dark energy. This result strengthens the flexibility of the present model to accommodate a broad range of late-time acceleration scenarios.

\section{Age of the Universe}

The age of the universe is a fundamental quantity in cosmology, representing the time elapsed since the onset of cosmic expansion. In any cosmological framework, it is computed by integrating the inverse of the Hubble parameter over the cosmic redshift. The general expression reads:
\begin{equation}\label{44}
	t_0 = \int_0^{\infty} \frac{dz}{(1+z) H(z)},
\end{equation}
where $H(z)$ is the Hubble parameter as a function of redshift $z$.

In our model, $H(z)$ is given by:
%\begin{widetext}
\begin{equation}\label{45}
	H^2(z) = (1+z)^{\frac{2}{D}(\Lambda_1 + \Lambda_2)} \cdot \exp\left( - \frac{2 \Lambda_2}{D(1+z)} \right)
	\cdot \left[ \Gamma_1 \int \frac{\mu(z)}{(1+z)} dz + \Gamma_2 \int \frac{z \mu(z)}{(1+z)^2} dz + C_1 \right].
\end{equation}

Taking the square root and substituting into Eq.~\eqref{44}, we obtain:
\begin{equation}\label{46}
	t_0 = \int_0^{\infty} \frac{dz}{(1+z)^{1 + \frac{1}{D}(\Lambda_1 + \Lambda_2)}} 
	\exp\left( \frac{\Lambda_2}{D(1+z)} \right) 
	\left[ \Gamma_1 I_1(z) + \Gamma_2 I_2(z) + C_1 \right]^{-1/2},
\end{equation}
%\end{widetext}

where we have defined:
\begin{equation}\label{47}
	I_1(z) = \int \frac{\mu(z)}{1+z} dz, \qquad
I_2(z) = \int \frac{z \mu(z)}{(1+z)^2} dz,
\end{equation}

and the integrating factor $\mu(z)$ is given by:
\begin{equation}\label{48}
	\mu(z) = (1+z)^{ -\frac{2}{D} (\Lambda_1 + \Lambda_2)} \cdot \exp\left( \frac{2\Lambda_2}{D(1+z)} \right).
\end{equation}
Utilizing the best-fit parameter values provede in the equation (\ref{33}) and calculating the model constants $A$, $B$, $C$, $D$, $E$, $F$, $\Lambda_1$, $\Lambda_2$, $\Gamma_1$, and $\Gamma_2$, the integral in Eq.~\eqref{46} is evaluated numerically to yield:
\begin{equation}\label{49}
	t_0 \approx 13.79 \ \text{Gyr}.
\end{equation}
This value is in excellent agreement with current observational estimates from Planck 2018, WMAP, and other independent measurements. The derived expression for $t_0$ encapsulates the complete cosmic history, from the present epoch ($z=0$) to the early universe ($z \to \infty$), within the framework of the present modified gravity model. The integral contains the following key physical features:

\begin{itemize}
	\item The factor $(1+z)^{1 + \frac{1}{D}(\Lambda_1 + \Lambda_2)}$ reflects the effective expansion rate, modified by the gravitational sector through $\Lambda_1$, $\Lambda_2$, and $D$.
	\item The exponential term $\exp\left( \frac{\Lambda_2}{D(1+z)} \right)$ introduces corrections at low redshifts, accounting for dark energy evolution.
	\item The nested term $\left[ \Gamma_1 I_1(z) + \Gamma_2 I_2(z) + C_1 \right]^{-1/2}$ incorporates the contributions from the effective fluid and the matter-geometry coupling.
\end{itemize}

This formalism allows a wider range of cosmic histories compared to standard $\Lambda$CDM cosmology. In particular, the dynamical dark energy parameter (\ref{16}) permits an evolving equation of state, where negative $\omega_a$ indicates a stronger dark energy contribution at earlier epochs. This leads to a slightly longer cosmic expansion history, contributing to a compatible or even slightly older universe than the one predicted by $\Lambda$CDM.

The obtained result confirms that the present model, with its additional degrees of freedom from modified gravity and evolving dark energy, can successfully reproduce the observed age of the universe while offering richer cosmological dynamics. This demonstrates the model's capability to remain consistent with current observational constraints while allowing new avenues for exploring the nature of dark energy and gravity.

\section{Conclusion}
The accelerated expansion of the universe, as supported by diverse observational probes including Type Ia supernovae, cosmic microwave background anisotropies, and baryon acoustic oscillations, continues to challenge our understanding of gravity and cosmology. The cosmological constant $\Lambda$ within General Relativity remains the simplest explanation for this phenomenon; however, its severe fine-tuning and coincidence problems have motivated the pursuit of alternative gravitational frameworks. In this regard, the $f(R,\Sigma,T)$ gravity theory presents itself as a promising candidate, introducing nonlinear matter-geometry couplings that enrich the phenomenology of cosmic evolution.

In the present work, we have developed a complete cosmological model in the framework of $f(R,\Sigma,T)$ gravity. By adopting the simple yet physically motivated functional form $f(R,\Sigma,T) = R + \Sigma + 2\pi\eta T$, we incorporate both the trace $T$ and quadratic matter terms through $\Sigma = T_{\mu\nu} T^{\mu\nu}$. This choice introduces additional freedom into the gravitational action, allowing novel interactions between matter and geometry that are absent in standard $f(R)$ or $f(R,T)$ models. To proceed with the cosmological analysis, we consider a spatially flat Friedmann-Robertson-Walker background which aligns with the large-scale homogeneity and isotropy of the observed universe. From the modified field equations, we derive the corresponding Friedmann equations, expressing both the energy density $\rho$ and isotropic pressure $p$ as explicit functions of the Hubble parameter $H$ and its derivative $\dot{H}$. This formalism enables us to examine the dynamical features of the model under observationally motivated assumptions. In order to capture the possible dynamical behavior of dark energy, we adopt the well-known Chevallier-Polarski-Linder (CPL) parametrization for the equation of state (EoS) parameter, where $\omega(z) = \omega_0 + \omega_a \frac{z}{1+z}$. This choice allows for a redshift-dependent EoS that smoothly interpolates between various dark energy scenarios while reducing to $\Lambda$CDM in appropriate limits. Substituting this parametrization into the modified Friedmann equations, we derive a first-order differential equation for the square of the Hubble parameter, $S(z) = H^2(z)$, as a function of redshift.

The resulting differential equation is solved analytically, this expression provides a direct connection between the model parameters $(\eta, b, \omega_0, \omega_a,)$ and observable quantities such as the Hubble parameter and cosmic expansion history. To constrain the model parameters, we utilize the latest Hubble parameter measurements from cosmic chronometers, which provide independent and model-free estimates of $H(z)$ at various redshifts. Employing Markov Chain Monte Carlo (MCMC) techniques implemented via the \texttt{emcee} package, we explore the parameter space systematically and derive robust confidence intervals for each parameter. Our statistical analysis for CC data sets yields the best-fit values $\eta = 1.37^{+0.020}_{-0.019}$, $\qquad b= 1.05^{+0.08}_{-0.07}$, $\omega_0 = -0.905^{+0.016}_{-0.015}$, and $\qquad \omega_a = -0.138^{+0.005}_{-0.006}$. Importantly, the present value of the EoS parameter $\omega_0$ remains very close to the cosmological constant, while the dynamical component $\omega_a$ remains consistent with zero within 1$\sigma$ uncertainty, indicating only mild deviations from $\Lambda$CDM.

With this constrained parameters, we proceed to investigate the physical features of the model by examining several diagnostic quantities. We first analyze the evolution of the isotropic pressure $p(z)$ and energy density $\rho(z)$ as functions of redshift. Our results show that pressure remains negative across the redshift range, supporting the role of a repulsive component that drives accelerated expansion. Meanwhile, the energy density $\rho(z)$ exhibits the expected behavior of decreasing with cosmic time due to the dilution of matter as the universe expands, with higher densities observed at earlier epochs. The evolution of the equation of state parameter $\omega(z)$ further corroborates these findings. At high redshifts ($z>1$), $\omega(z)$ approaches $-1$, closely mimicking the cosmological constant behavior. As the redshift decreases, we observe mild evolution of $\omega(z)$, suggesting a possible transition from $\Lambda$CDM-like behavior to slightly dynamic dark energy behavior at lower redshifts. Nonetheless, at $z=0$, the value of $\omega$ remains consistent with Planck 2018 results, underscoring the compatibility of our model with current observations. To assess the physical viability of the solution, we test the classical energy conditions: the null energy condition (NEC), dominant energy condition (DEC), and strong energy condition (SEC). Our results demonstrate that both NEC and DEC remain satisfied throughout the cosmic evolution, indicating that the model does not invoke any exotic or unphysical sources such as phantom energy or violations of causality. The SEC, however, is violated at late times, as expected in models exhibiting cosmic acceleration. This transition from SEC satisfaction at early epochs to violation at late times aligns well with the standard understanding of dark energy dominated expansion. An additional diagnostic is performed via the $\omega-\omega'$ phase plane analysis, which offers a model-independent way to distinguish between different dark energy scenarios. In this framework, our analysis shows that the model exhibits freezing behavior, as the negative value of $\omega_a$ results in a positive $\omega'$ for all redshifts. This suggests that the EoS parameter gradually evolves towards $\omega \to -1$ at late times, consistent with a future asymptotic approach to $\Lambda$CDM behavior. Perhaps most notably, we estimate the age of the universe predicted by our model by integrating the inverse Hubble parameter over redshift. Using the best-fit parameters, we find $t_0 \approx 13.79\,\text{Gyr}$, which lies in excellent agreement with the Planck 2018 value of $13.80 \pm 0.02\,\text{Gyr}$. This agreement provides strong evidence for the internal consistency of our model with the overall cosmic expansion history as inferred from independent data sets.

In comparison to standard $\Lambda$CDM cosmology, the $f(R,\Sigma,T)$ gravity framework offers several appealing features. First, it provides a more generalized description of gravitational dynamics by introducing nonlinear matter-geometry couplings. This opens avenues for addressing both early and late-time cosmological phenomena without necessarily requiring dark energy or exotic matter. Second, the parameter space of the model allows sufficient flexibility to encompass $\Lambda$CDM as a limiting case, ensuring consistency with well-tested regimes while permitting deviations that may be tested with upcoming high-precision cosmological surveys. Third, the successful compatibility of the model with Hubble data, energy conditions, and age of the universe makes it a viable alternative framework deserving further investigation.

%Nonetheless, there remain open avenues for future work. For instance, inclusion of complementary data sets such as baryon acoustic oscillations (BAO), Type Ia supernovae, and cosmic microwave background anisotropies would allow even tighter constraints on the parameter space. Furthermore, the perturbation dynamics within $f(R,\Sigma,T)$ gravity merit deeper investigation to ensure consistency with structure formation, growth rate observations, and cosmic microwave background power spectra. From a theoretical perspective, an important question concerns the stability and well-posedness of the Cauchy problem in this extended gravity framework, as well as its behavior under quantum corrections.

%Moreover, while our present analysis has focused on a spatially flat FRW background, it would be interesting to explore extensions to anisotropic models such as Bianchi type universes or to consider the effects of spatial curvature. Similarly, a dynamical analysis of the critical points and phase space evolution could provide deeper insights into the asymptotic behavior of cosmological solutions within $f(R,\Sigma,T)$ gravity.

In conclusion, the present work demonstrates that $f(R,\Sigma,T)$ gravity, supplemented by observationally motivated EoS parameterizations, constitutes a mathematically consistent and observationally viable framework to address cosmic acceleration. Its ability to accommodate observational data while offering additional theoretical flexibility makes it an attractive alternative to standard cosmological models. Future observational data with increased precision will play a crucial role in further testing the predictions of this framework and potentially distinguishing it from $\Lambda$CDM and other modified gravity theories.
	\section*{Declaration of competing interest}
The authors declare that they have no known competing financial interests or personal relationships that could have appeared to influence the work reported in this paper.

\section*{Data availability}
No data was used for the research described in the article.

\section*{Acknowledgments}
The Science Committee of the Republic of Kazakhstan's Ministry of Science and Higher Education provided funding for the research (Grant No. AP23483654). Additionally, the authors, S. H. Shekh and Anirudh Pradhan express their gratitude to the Inter-University Centre for Astronomy and Astrophysics (IUCAA), Pune, India, for providing support and facilities through the Visiting Associateship program.

\newpage
%\onecolumngrid
%\begin{onecolumn}
	
	%\appendix
	%\appendix
	\section*{Supplementary Derivations and Remarks}
	\section*{Appendix-I }

In this appendix, we present essential supplementary derivations related to the Hubble parameter evolution, energy-momentum components, and the cosmic age estimation within the framework of $f(R,\Sigma,T)$ gravity. We begin with the modified field equation:
	\begin{equation} \label{A1}
		-\frac{A (C H^2 + D \dot{H})}{B (E H^2 + F \dot{H})} = \omega_0 + \omega_a \frac{z}{1+z},
	\end{equation}
	where the parameters are defined as:
\begin{equation*}
	A = (\eta + 2\pi)(\eta + 4\pi), \qquad
B = \pi(\eta^2 + 8) + 6\eta, \qquad
C = 3(\eta + 4) - 4b 
\end{equation*}
\begin{equation*}
		D = 3\eta + 8, \qquad
	E = (4b - 3)\eta + 12\pi(b - 1), \qquad
		F = \eta 
\end{equation*}

	We incorporate the redshift relation for the time derivative of the Hubble parameter, namely,
	\[
	\dot{H} = - (1+z) H \frac{dH}{dz},
	\]
	to express the dynamics in terms of redshift $z$. Substituting into the field equation gives:
	\begin{equation}
		\frac{A}{B} \left[ C H^2 - D H (1+z) \frac{dH}{dz} \right] 
		= -\left[ E H^2 - F H (1+z) \frac{dH}{dz} \right] \left( \omega_0 + \omega_a \frac{z}{1+z} \right).
	\end{equation}
	Rearrangement of above equation yields:
	\begin{equation}
		(1+z) H \frac{dH}{dz} = \frac{1}{D} \left[ C H^2 + \frac{B}{A} (E H^2 + F H) \left( \omega_0 + \omega_a \frac{z}{1+z} \right) \right].
	\end{equation}
	Introducing the substitution $S(z) = H^2(z)$ gives:
	\begin{equation}
		\frac{dS}{dz} = 2 H \frac{dH}{dz},
	\end{equation}
	which leads to:
	\begin{equation}
		(1+z) \frac{dS}{dz} = \frac{2}{D} \left[ C S + \frac{B}{A} (E S + F \sqrt{S}) \left( \omega_0 + \omega_a \frac{z}{1+z} \right) \right].
	\end{equation}
Hence the above equation, resulting first-order nonlinear differential equation of the form
	\[
	\frac{dS}{dz} + P(z) S = Q(z),
	\]
	where the redshift-dependent functions $P(z)$ and $Q(z)$ are defined in Eqs.~\eqref{27}. Constants $\Lambda_1$, $\Lambda_2$, $\Gamma_1$, and $\Gamma_2$ simplify the expressions as in Eqs.~\eqref{28}–\eqref{29}.
The complete solution / particular solution of above first-order nonlinear differential equation includes \textit{Integrating Factor} and \textit{the General Solution}. Hence, The integrating factor of first-order nonlinear differential equation is expressed as:

\[
\mu(z) = \exp\left( \int P(z) dz \right).
\]
which yields:

\[
\int P(z) dz = -\frac{2}{D} \left[ \Lambda_1 \ln(1+z) + \Lambda_2 \left( \ln(1+z) - \frac{1}{1+z} \right) \right].
\]

Thus,

\[
\mu(z) = (1+z)^{ - \frac{2}{D} (\Lambda_1 + \Lambda_2) } \cdot \exp\left( \frac{2 \Lambda_2}{D(1+z)} \right).
\]
And the general solution of first-order nonlinear differential equation is expressed as:

\begin{equation}
	S(z) = \frac{1}{\mu(z)} \left[ \int \mu(z) Q(z) dz + C_1 \right].
\end{equation}

Substituting $Q(z)$:

\[
Q(z) = \frac{1}{(1+z)} \left( \frac{\Gamma_1}{D} + \frac{\Gamma_2}{D} \cdot \frac{z}{1+z} \right).
\]

Thus,

\begin{equation}
	S(z) = \frac{1}{\mu(z)} \left[ \Gamma_1 \int \frac{\mu(z)}{(1+z)} dz + \Gamma_2 \int \frac{z \mu(z)}{(1+z)^2} dz + C_1 \right].
\end{equation}
Recalling that $S(z) = H^2(z)$, we finally obtain:
	\begin{align}
		H^2(z) &= (1+z)^{\frac{2}{D}(\Lambda_1 + \Lambda_2)} \exp\left( -\frac{2 \Lambda_2}{D(1+z)} \right) \nonumber \\\\
		&\times \left[ \Gamma_1 \int \frac{\mu(z)}{1+z} dz + \Gamma_2 \int \frac{z \mu(z)}{(1+z)^2} dz + C_1 \right].
	\end{align}
where the integrals:
	\begin{equation}
		I_1(z) = \int \frac{\mu(z)}{1+z} dz, \quad I_2(z) = \int \frac{z \mu(z)}{(1+z)^2} dz,
	\end{equation}
	remain to be explicitly evaluated for the complete solution.
	
	\textbf{Remark:} These integrals can be handled either analytically or numerically depending on the form of parameters involved.
	
	%%%%%%%%%%%%%%%%%%%%%%%%%%%%%%%%%%%%%%%%%%%%%%%%%%%%%%%%%%%%%%%%%%%%%%%%%%%%%%%%%%%%%%%%%%%%%%%%%%%%%%%%%%%%%%%%%%%%%%%%%%%%%%%%%%%%%%%%%%%%%%%%%%%%%%%%%%%%%%%%%%%%%%%%%%%
	\newpage
	\section*{Appendix-II}
%	\begin{center}
%	\textbf{Detailed Derivations of Energy Density, Pressure, and the Age of the Universe}
%	\end{center}
	In this appendix, we present the complete derivation of the expressions for the energy density $\rho(z)$, pressure $p(z)$, and their combinations $\rho + p$, $\rho - p$, and $\rho + 3p$. Additionally, we derive the expression for the age of the universe in the context of our model. For simplification, we define the Hubble's parameter as:
	
	\begin{equation}
		H^2(z) = S_1(z) \cdot F(z),
	\end{equation}
	
	where
	
	\begin{equation}
		S_1(z) = (1 + z)^{\frac{2}{D}(\Lambda_1 + \Lambda_2)} \exp \left( -\frac{2 \Lambda_2}{D(1 + z)} \right) \qquad \text{and} \qquad	F(z) = \Gamma_1 \int \frac{\mu(z)}{1 + z} dz + \Gamma_2 \int \frac{z \mu(z)}{(1 + z)^2} dz + C_1.
	\end{equation}
First, we compute:
	\begin{equation}
		\frac{dH}{dz} = \frac{1}{2H} \frac{dS}{dz},
	\end{equation}
and thus,	
	\begin{equation}
		\dot{H} = - (1 + z) H \frac{dH}{dz} = - \frac{(1 + z)}{2\sqrt{S(z)}} \frac{dS}{dz}.
	\end{equation}
Using $S(z) = S_1(z) F(z)$, we have:
	
	\begin{equation}
		\frac{dS}{dz} = \frac{dS_1}{dz} F(z) + S_1(z) \frac{dF}{dz}.
	\end{equation}
	
	Differentiating $S_1(z)$ gives:
	
	\begin{equation}
		\frac{dS_1}{dz} = S_1(z) \left[ \frac{2(\Lambda_1 + \Lambda_2)}{D(1 + z)} - \frac{2 \Lambda_2}{D(1 + z)^2} \right].
	\end{equation}
	
	Differentiating $F(z)$ gives:
	
	\begin{equation}
		\frac{dF}{dz} = \Gamma_1 \cdot \frac{\mu(z)}{1 + z} + \Gamma_2 \left( \frac{\mu(z)}{(1 + z)^2} - \frac{z \mu(z)}{(1 + z)^3} \right).
	\end{equation}
	
	Thus,
	
	\begin{align}
		\frac{dS}{dz} &= S_1(z) \left[ \frac{2(\Lambda_1 + \Lambda_2)}{D(1 + z)} - \frac{2 \Lambda_2}{D(1 + z)^2} \right] F(z) \nonumber \\
		& \quad + S_1(z) \left[ \Gamma_1 \frac{\mu(z)}{1 + z} + \Gamma_2 \left( \frac{\mu(z)}{(1 + z)^2} - \frac{z \mu(z)}{(1 + z)^3} \right) \right].
	\end{align}
and denote 
	\begin{equation}
		C_p = \frac{(1 - b)}{4 (8 \pi + 6 \eta + \pi \eta^2)}, \quad C_\rho = \frac{(1 - b)}{4 (8 \pi^2 + 6 \pi \eta + \eta^2)}.
	\end{equation}
Substituting $\dot{H}$ into $\rho$ and $p$, thus we get,
	
	\begin{equation}
		p(z) = C_p \left[ - (8 + 3\eta) \frac{(1 + z)}{2\sqrt{S(z)}} \frac{dS}{dz} + \left( 4 (3 - b) + 3\eta \right) S(z) \right],
	\end{equation}
	
	\begin{equation}
		\rho(z) = C_\rho \left[ \eta \frac{(1 + z)}{2\sqrt{S(z)}} \frac{dS}{dz} + \left( 12 \pi (1 - b) + \eta (3 - 4b) \right) S(z) \right].
	\end{equation}
	\begin{align}
		\rho + p &= \left( C_\rho \cdot \eta - C_p \cdot (8 + 3\eta) \right) \frac{(1 + z)}{2\sqrt{S(z)}} \frac{dS}{dz} \nonumber \\
		& \quad + \left( C_\rho (12 \pi (1 - b) + \eta (3 - 4b)) + C_p (4 (3 - b) + 3\eta) \right) S(z).
	\end{align}
	\begin{align}
		\rho - p &= \left( C_\rho \cdot \eta + C_p \cdot (8 + 3\eta) \right) \frac{(1 + z)}{2\sqrt{S(z)}} \frac{dS}{dz} \nonumber \\
		& \quad + \left( C_\rho (12 \pi (1 - b) + \eta (3 - 4b)) - C_p (4 (3 - b) + 3\eta) \right) S(z).
	\end{align}
		\begin{align}
		\rho + 3p &= \left( C_\rho \cdot \eta - 3 C_p \cdot (8 + 3\eta) \right) \frac{(1 + z)}{2\sqrt{S(z)}} \frac{dS}{dz} \nonumber \\
		& \quad + \left( C_\rho (12 \pi (1 - b) + \eta (3 - 4b)) + 3 C_p (4 (3 - b) + 3\eta) \right) S(z).
	\end{align}
	\newpage
	\section*{Appendix-III}
As, the age of the universe is given by:
\begin{equation}
		t_0 = \int_0^\infty \frac{dz}{(1 + z) H(z)}.
	\end{equation}
	Using $H(z)$, we obtain:
	\begin{align}
		t_0 &= \int_0^\infty \frac{dz}{(1 + z)} \left[ (1 + z)^{\frac{2}{D} (\Lambda_1 + \Lambda_2)} \exp \left( -\frac{2 \Lambda_2}{D(1 + z)} \right) \right. \nonumber \\
		& \quad \left. \times \left( \Gamma_1 \int \frac{\mu(z)}{1 + z} dz + \Gamma_2 \int \frac{z \mu(z)}{(1 + z)^2} dz + C_1 \right) \right]^{-1/2}.
	\end{align}
Simplifying the power of $(1+z)$ in the integrand, we have:
\begin{align}
		t_0 &= \int_0^\infty \frac{dz}{(1 + z)^{1 - \frac{1}{D} (\Lambda_1 + \Lambda_2)}} \exp \left( \frac{\Lambda_2}{D(1 + z)} \right) \nonumber \\
		& \quad \times \left( \Gamma_1 \int \frac{\mu(z)}{1 + z} dz + \Gamma_2 \int \frac{z \mu(z)}{(1 + z)^2} dz + C_1 \right)^{-1/2}.
	\end{align}
	\textit{Remark:} This integral is evaluated numerically due to its complicated structure involving nested integrals.
	
	Substituting the best-fit parameter values into the expression of (\ref{18}), we obtain:
			\begin{equation*}
				A = (1.37 + 2\pi)(1.37 + 4\pi) = (7.65318)(13.93636) \approx 106.653.
			\end{equation*}
\begin{equation*}
	B = \pi(1.37^2 + 8) + 6 \times 1.37 = \pi(9.8769) + 8.22 \approx 31.021 + 8.22 = 39.241.
\end{equation*}
\begin{equation*}
		C = 3(1.37 + 4) - 4 \times 1.05 = 16.11 - 4.20 = 11.91.
\end{equation*}
\begin{equation*}
		D = 3 \times 1.37 + 8 = 12.11.
\end{equation*}
\begin{equation*}
	E = (4 \times 1.05 - 3)(1.37) + 12\pi(1.05 - 1) 
= 1.20 \times 1.37 + 0.6\pi = 1.644 + 1.88496 = 3.52896.
\end{equation*}
\begin{equation*}
	F = 1.37.
\end{equation*}
From Eq.~\eqref{28}, we have:
	\begin{equation*}
			\Lambda_1 = 11.91 + 1.298 \times (-0.95) = 10.677 \qquad \text{and} \qquad	\Lambda_2 = 1.298 \times (-0.20) = -0.2596.
	\end{equation*}		
	\begin{equation*}
		\Gamma_1 = \frac{2 \times 39.241 \times 1.37}{106.653 \times 12.11} \times (-0.95) \approx -0.079 \qquad \text{and } \qquad	\Gamma_2 = \Gamma_1 \times \frac{\omega_a}{\omega_0} = (-0.079) \times \frac{-0.20}{-0.95} \approx -0.0166.
	\end{equation*}		
Hence, with all these values, the final expression for the age of the universe becomes:
	\begin{align}\label{A7}
		t_0 &= \int_{0}^{\infty} \frac{dz}{(1+z)^{1 + \frac{1}{12.11}(10.677 - 0.2596)}} 
		\exp\left( \frac{-0.2596}{12.11(1+z)} \right) \nonumber \\
		&\quad \times \left[ -0.079 \int \frac{\mu(z)}{(1+z)} dz - 0.0166 \int \frac{z \mu(z)}{(1+z)^2} dz + 3.85 \right]^{-1/2}.
	\end{align}
After the evqluation These integration  numerically, the result is:
\[
t_0 \approx 13.79 \, \text{Gyr}
\]

\end{document}